\documentclass[preprint,12pt, a4paper]{elsarticle}

\usepackage{amssymb}

\usepackage{lineno}
\usepackage{booktabs}
\usepackage{float}

\journal{SoftwareX}

\usepackage{hyperref}
\usepackage{color,soul}
\usepackage{xspace}         
\usepackage{microtype}      
\usepackage{wrapfig}        
\usepackage[htt]{hyphenat}

\restylefloat{table}

\newcommand{\coo}{\ensuremath{\mathrm{CO_2}}}

\newcommand{\DIETERpy}{DIETERpy\xspace}
\newcommand{\DIETERgms}{DIETERgms\xspace}

\usepackage{forest}

\definecolor{folderbg}{RGB}{124,166,198}
\definecolor{folderborder}{RGB}{110,144,169}

\def\Size{4pt}
\tikzset{
      folder/.pic={
        \filldraw[draw=folderborder,top color=folderbg!50,bottom color=folderbg]
          (-1.05*\Size,0.2\Size+5pt) rectangle ++(.75*\Size,-0.2\Size-5pt);  
        \filldraw[draw=folderborder,top color=folderbg!50,bottom color=folderbg]
          (-1.15*\Size,-\Size) rectangle (1.15*\Size,\Size);
      }
    }

\usepackage{anysize}
\marginsize{2.5cm}{2.5cm}{1.5cm}{2cm}
\usepackage[bottom]{footmisc}

\begin{document}

\begin{frontmatter}



\title{DIETERpy: a Python framework for The Dispatch and Investment Evaluation Tool with Endogenous Renewables}


\author{Carlos~Gaete-Morales\fnref{myfootnote}}
\fntext[myfootnote]{Corresponding author}
\ead{cgaete@diw.de}

\author{Martin~Kittel}
\ead{mkittel@diw.de}

\author{Alexander~Roth}
\ead{aroth@diw.de}

\author{Wolf-Peter~Schill}
\ead{wschill@diw.de}

\address{All authors are at the German Institute for Economic Research (DIW Berlin), Mohrenstr. 58, D-10117 Berlin, Germany\vspace{-3\baselineskip}}

\begin{abstract}

\noindent \small DIETER is an open-source power sector model designed to analyze future settings with very high shares of variable renewable energy sources. It minimizes overall system costs, including fixed and variable costs of various generation, flexibility and sector coupling options. Here we introduce \DIETERpy that builds on the existing model version, written in the General Algebraic Modeling System (GAMS), and enhances it with a Python framework. This combines the flexibility of Python regarding pre- and post-processing of data with a straightforward algebraic formulation in GAMS and the use of efficient solvers. \DIETERpy also offers a browser-based graphical user interface. The new framework is designed to be easily accessible as it enables users to run the model, alter its configuration, and define numerous scenarios without a deeper knowledge of GAMS. Code, data, and manuals are available in public repositories under permissive licenses for transparency and reproducibility.
\end{abstract}


\end{frontmatter}



\newpage

\section*{Code metadata} \label{sec:Code metadata}

\begin{table}[H]
\resizebox{\textwidth}{!}{
\begin{tabular}{l l}
\hline
Current code version & v0.3.0 \\
Permanent link to code/repository used for this code version & \url{https://gitlab.com/diw-evu/dieter_public/dieterpy} \\
Legal Code License & MIT \\
Code versioning system used & git \\
Software code languages, tools, and services used & Python, GAMS \\
Compilation requirements, operating environments \& dependencies & Python 3.6, GAMS +24.8 \\
Link to developer documentation/manual & \url{https://diw-evu.gitlab.io/dieter_public/dieterpy} \\
Support email for questions & Carlos Gaete \href{mailto:cgaete@diw.de}{cgaete@diw.de} \\
\hline
\end{tabular}
}
\end{table}

\section*{Software metadata} \label{sec:Software metadata}
\begin{table}[H]
\resizebox{\textwidth}{!}{
\begin{tabular}{l l}
\hline
Current software version & v0.3.0 \\
Permanent link to executables of this version & \url{https://pypi.org/project/dieterpy/0.3.0/} \\
Legal Code License & MIT \\
Computing platform/Operating Systems & Linux, Microsoft Windows \\
Installation requirements & Python 3.6, GAMS +24.8, conda  \\
Link to developer documentation/manual & \url{https://diw-evu.gitlab.io/dieter_public/dieterpy} \\
Support email for questions & Carlos Gaete \href{mailto:cgaete@diw.de}{cgaete@diw.de} \\
\hline
\end{tabular}
}
\end{table}

\setcounter{table}{0}



\section{Motivation and Significance}






Mitigating climate change calls for a decarbonization of economies around the world. Power sectors are among the most \coo-intensive sectors, and renewable energy sources play a major role in decarbonizing them \citep{IPCC_2018}. Wind and solar energy is abundant, but come with specific characteristics, such as high fixed and low variable costs, and temporally variable production patterns. Their cost-efficient large-scale integration into energy systems calls for numerical analyses. As to how they can replace fossil fuels and become the backbone of future energy systems particularly raises questions on the role of temporal flexibility options and sector coupling.
    
DIETER is a power sector model developed to address such questions of temporal flexibility options and sector coupling in future scenarios with high shares of variable renewable energy. While the model was first used to investigate the role of electricity storage, it has subsequently been extended to also consider various sector coupling options, such as battery-electric vehicles (BEV), power-to-heat, green hydrogen, as well as solar prosumage. 

Here we introduce a new major development of the model, \DIETERpy. Its core model code builds on previous versions of DIETER, written in the General Algebraic Modeling System (GAMS), to which we refer as \DIETERgms from now on. DIETER has been first introduced to the literature in the context of analyzing long-term electricity storage needs for variable renewable energy sources \cite{ZERRAHN_2017, SCHILL_2018}. The functionalities added and enhanced by \DIETERpy provide new tools for simpler and more comprehensive scenario runs, facilitate a more convenient configuration of the model, and enhance accessibility for users. In \DIETERpy, the original algebraic GAMS model is embedded in a Python framework, or wrapper, that is responsible for the (1) configuration of the model, (2) definition of the scenarios to be investigated, and (3) pre- and post-processing of the data. 

Python, as an interpreted language, has been used as a wrapper for many highly efficient programs in their respective fields, such as C/C++, Fortran, or GAMS. In the literature there are several examples of applications in energy and climate change, for instance, the Python interface for the Dutch Atmospheric Large Eddy Simulation, or the Python Wrapper for System Advisor Model SAM \cite{VANDENOORD_2020, PySAM_2019}. Other examples for GAMS models with Python wrappers in the energy modeling space include IIASA's MESSAGEix modeling framework and the dispatch model Dispa-SET \citep{HUPPMANN_2019, Kavvadias_2018}. To our knowledge, \DIETERpy is the first capacity expansion power sector model that uses the GAMS API for Python. We also go one step further than previous approaches of Python model wrappers by developing a tool that enables the creation of scenarios by modifying parameter values, variable bounds and (de-)activation of constraints without the need for altering the original model code. \DIETERpy also allows running model instances in parallel by using several processor cores and the implementation of the GUSS tool, an advanced GAMS feature that reduces the compilation times of similar scenarios through model instances. Not least, embedding our legacy model DIETER in a Python framework substantially enhances the functionalities of the previous pure GAMS-based version, which may be useful for many current and future users of the model.

\DIETERpy can easily be installed via Python package managers, and no extensive knowledge of GAMS is needed for initial runs. The model can be configured either by a browser-based graphical user interface or by \texttt{CSV} files. For standard scenario runs, the GAMS-based core model code does not have to be altered by the user, but only for more fundamental model changes. \DIETERpy has a data post-processing routine that collects the results of multiple model runs and makes them accessible for further analysis in different output formats, allowing users to proceed with their tools of choice. Basic model results can also be visualized within the browser interface.

\section{Software Description}




DIETER is a power sector capacity expansion model based on linear equations. It minimizes total system costs in a long-run equilibrium setting under perfect foresight. In economic terms, it takes a social planner perspective. Its objective function is the sum of investment costs into various generation and flexibility technologies as well as transmission, and related variable costs for a given time period (usually a full year). DIETER aims to minimize these costs, given exogenous techno-economic input parameters and time series such as demand and renewable availability which are provided in an hourly resolution. Hence, the solution is an optimal portfolio of generation and flexibility technologies such as electricity storage, as well as their optimal dispatch for the given time period. The model is solved for all consecutive hours of a target year. 

A range of equations implements constraints with respect to generation capacities, renewable shares, \coo~emissions, renewable availability, and balancing reserves. Further, there are inter-temporal restrictions related to various types of storage and sector coupling. The model does not include a detailed representation of the underlying transmission grid infrastructure. Within regions, e.g.,~European countries, it makes a copper-plate assumption; between regions, DIETER uses a simple transport model approach with Net Transfer Capacities (NTCs). Different versions of DIETER have been developed and applied in several research projects, which has led to a growing number of publications in relevant field journals and several ongoing working papers (see Section~\ref{sec:impact}). 

From the beginning, DIETER was designed as a lean, tractable and flexible open-source tool. The model applications that were published so far hardly required high-performance computing resources, but could largely be solved on a standard computer. The new \DIETERpy framework aims to further increase the model's accessibility and usefulness also for casual users and practitioners. 

\DIETERpy includes some basic data in its installation, which the user may have to adjust, depending on the types of scenario analyses to be investigated. We provide a range of techno-economic input parameters for the available technology portfolio. This includes time-invariant costs parameters related to investment in various generation and flexibility technologies as well as cross-border transfer capacities, fuel costs, flexibility costs, and variable costs of different technologies, as well as technology-specific efficiencies, technical life-times, and upper/lower capacity expansion bounds. Further, we include geographical characteristics, as well as time series of renewable availability and demand for electricity and heat. All of the data is freely available and sources are listed within the respective spreadsheets.

\subsection{Software Architecture} \label{sec:software architecture}



\begin{figure}[b!]
    \centering
    \includegraphics[width=1.0\textwidth]{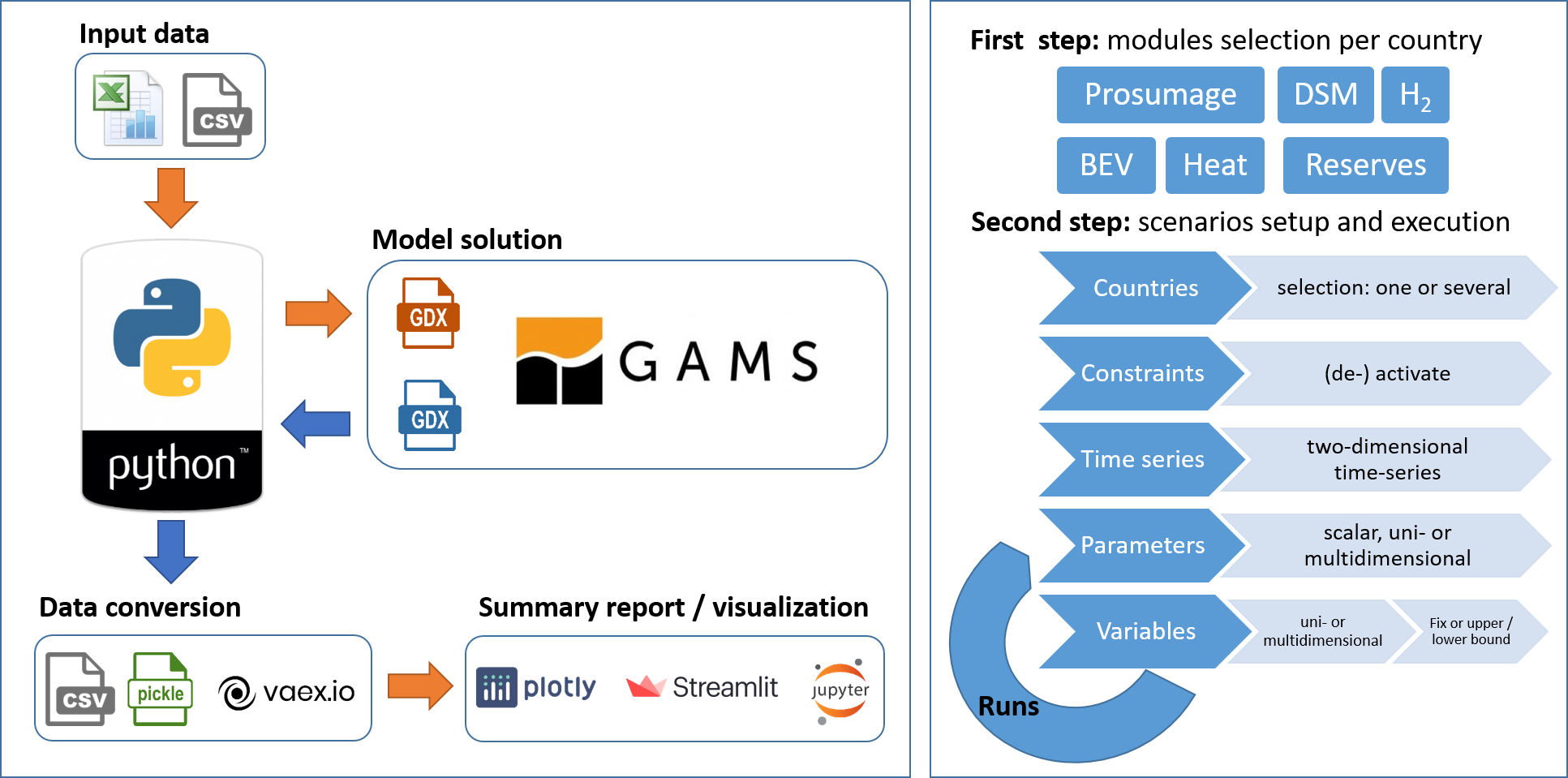}
    \caption{Graphical overview of \DIETERpy (source: own illustration).}
    \label{fig:overview}
\end{figure}

\DIETERpy is a software package written in the programming languages Python and GAMS. It can be installed via the Python package manager \textit{pip}. Running the model requires a license for the GAMS Base Module and an LP solver (we use CPLEX).\footnote{The GAMS Software GmbH told us that potential users who want to explore \DIETERpy can ask for a test license, \href{mailto:sales@gams.com}{sales@gams.com}.} The left panel of Figure~\ref{fig:overview} provides an overview of the software architecture of \DIETERpy. Data pre- and post-processing as well as the management of different scenario runs are carried out with Python, only the optimization itself is done within a GAMS module that is initiated via Python. The user can choose between different data output formats to further analyze the results.

The right panel of Figure~\ref{fig:overview} summarizes the two steps required before running optimization with \DIETERpy. The first step consists of selecting modules. The basic module contains all variables and constraints for a default power sector dispatch and investment model. Additional model features, such as demand-side management (\texttt{dsm}), battery-electric vehicles (features \texttt{ev\_endogenous} and \texttt{ev\_exogenous}), balancing reserves (\texttt{reserves}), solar prosumage (\texttt{prosumage}) and power-to-heat (\texttt{heat}), can be enabled. In a second step, multiple scenario runs can be specified. Within a scenario run, it is possible to modify parameters, time series and variable bounds, and select different constraints and sets of countries. Parameters and variables have to be indicated with their domain unless it is a scalar or a dimensionless variable. The domain must contain set names or set elements. Unidimensional and multidimensional parameters and variables are supported. For variables, we can establish upper and lower bounds, as well as fixed values.

\subsection{Software Functionalities} \label{sec:software functionalities}

\begin{wrapfigure}{R}{.5\linewidth}
\centering 
{\footnotesize
\begin{forest}
      for tree={
        font=\sffamily,
        grow'=0,
        child anchor=west,
        parent anchor=south,
        anchor=west,
        calign=first,
        inner ysep=0pt,
        inner xsep=8pt,
        edge path={
          \noexpand\path [draw, \forestoption{edge}]
          (!u.south west) +(7.5pt,0) |- (.child anchor) pic {folder} \forestoption{edge label};
        },
        file/.style={edge path={\noexpand\path [draw, \forestoption{edge}]
          (!u.south west) +(7.5pt,0) |- (.child anchor) \forestoption{edge label};},
          inner xsep=2pt,font=\small\sffamily
                     },
        before typesetting nodes={
          if n=1
            {insert before={[,phantom]}}
            {}
        },
        fit=band,
        before computing xy={l=15pt},
      }  
    [[Project name
     [manage.py,file]
     [projectfiles
      [data\_input
       [static\_input.xlsx,file]
       [timeseries\_input.xlsx,file]
      ]
      [iterationfiles
       [iteration\_table.csv,file]
       [iteration\_data.xlsx,file]
      ]
      [model
       [model.gms,file]
      ]
      [settings
       [constraints\_list.csv,file]
       [features\_node\_selection.csv,file]
       [reporting\_symbols.csv,file]
       [project\_variables.csv,file]
      ]
     ]
    ]]
\end{forest}
}
\caption{Folders and files tree of a project (source: own illustration).}\label{fig:projecttree}
\end{wrapfigure}
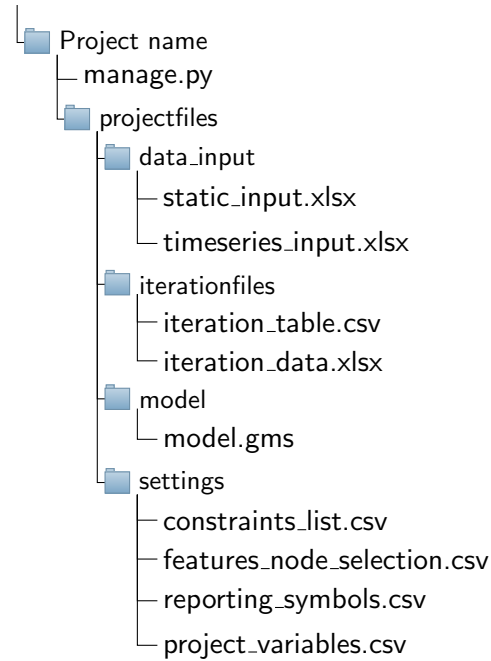

In \DIETERpy, the user can select modules and other settings using an easy-to-access browser-based graphical user interface, or editing the corresponding configuration \texttt{CSV} files. In order to get access to the configuration files, a project has to be created. This can be done by calling the \texttt{create\_project} function after typing \texttt{dieterpy} in the command line. This will result in a new folder that entails all configuration files and input data. Figure~\ref{fig:projecttree} shows the folders and the file tree of a project.

Out-of-the box, we allow the user to alter: 
\begin{samepage}
\begin{enumerate}
    \item basic program settings (\texttt{project\_variables.csv}), 
    \item modules (\texttt{features\_node\_selection.csv}), and 
    \item the scenario table (\texttt{iteration\_table.csv}). 
\end{enumerate}
\end{samepage}

Basic program settings include, for instance, the definition of input files, and the output file formats; here, the user can choose between \texttt{CSV}, \texttt{Pickle}, and \texttt{VAEX}. As \DIETERpy can solve scenarios in parallel, we also allow to adapt specific settings concerning parallelization. In the ``basic program settings'', the user may further switch between an ``investment \& dispatch model'' and a ``dispatch only model''. Cross-border electricity flows can also be easily switched on and off (see Table~\ref{tab:project variables}).

\begin{wraptable}{R}{.60\linewidth}
\small
\centering
\caption{Example of the project\_variables.csv file (source: own illustration).}
\label{tab:project variables}
\begin{tabular}{ll}
\toprule
Variable                        &                 Value \\
\midrule
scenarios\_iteration            &                   yes \\
skip\_input                     &                    no \\
skip\_iteration\_data\_file     &                    no \\
base\_year                      &                  2030 \\
end\_hour                       &                 h8760 \\
dispatch\_only                  &                    no \\
network\_transfer               &                   yes \\
no\_crossover                   &                   yes \\
infeasibility                   &                    no \\
GUSS                            &                   yes \\
GUSS\_parallel                  &                   yes \\
GUSS\_parallel\_threads         &                     0 \\
data\_input\_file               &      static\_input.xlsx \\
time\_series\_file              &     timeseries\_input.xlsx \\
iteration\_data\_file           &  iteration\_data.xlsx \\
gdx\_convert\_parallel\_threads &                     0 \\
gdx\_convert\_to\_csv           &                    no \\
gdx\_convert\_to\_pickle        &                   yes \\
gdx\_convert\_to\_vaex          &                    no \\
report\_data                    &                   yes \\
\bottomrule
\end{tabular}
\end{wraptable}

\DIETERpy is a generic model which can be adapted to any geographic setting. In its default version, the calibration is most refined for Germany, yet eleven other countries (France, Denmark, Belgium, Netherlands, Poland, Czech Republic, Austria, Switzerland, Spain, Italy, and Portugal) can also be activated. Adding more countries or applying the model to a different geographic region can be done relatively easily if respective input data on electricity demand as well as meaningful bounds for generation and transfer capacity are available. Different modules can be switched on for different countries, depending on the research question and model size restrictions, as shown in Table~\ref{tab:Features_node}.\footnote{By the time of writing, input data is still missing for several country-module combinations. Further, a green hydrogen module is also available \citep{stoeckl2020}, but only in the pure GAMS version (\DIETERgms).}

\DIETERpy further provides an easy way to change single parameter values or define and control entire scenario runs. The program has an easy-to-modify scenario table (\texttt{iteration\_table.csv}) that allows the user to edit (a) the set of countries, (b) specific constraints specifications, (c) entire time series (such as demand or renewable capacity factors), (d) single parameter values and (e) variable bounds (either fixing it or providing upper or lower bounds). The scenario table is an easy-to-use, transparent tool for quickly running the model multiple times in different specifications, where every row in that table refers to one scenario (see Table~\ref{tab:main iteration}). The parametrization is established by adding column headings to the table that are dedicated to special features. To modify input parameters, we add a column heading that matches the parameter name in the \texttt{model.gms} file. For variables, we can set fixed values as well as upper and lower bounds.

Once the configuration files have been edited and adapted for a desired case study, the optimization is ready to start. The excel files that contain the input data and the configuration \texttt{CSV} files are loaded, and then a Python routine creates the optimization- and GAMS-compatible \texttt{GDX} files. Via the Python-API, the model (\texttt{model.gms}) as well as scenario table is passed to GAMS.

For solving, \DIETERpy uses the Python-API of GAMS to build a model instance. This model instance exists until a solver in GAMS returns the solutions. We have developed three options to run optimization problems with \DIETERpy. One option consists of running the scenarios with independent model instances that compile every time they are built. With this method, the scenarios are solved sequentially. This may lead to long execution times when several scenarios are solved due to the compilation time incurred for each model instance. This problem is solved in part with the second option, the Gather-Update-Solve-Scatter (GUSS) tool in GAMS. It allows running several scenarios with only one model instance by updating the variables, parameters and equation values from one scenario to another. The third option takes advantage of multiprocessing in Python. It enables to run the GUSS tool in several processor cores, in which several scenarios are solved in parallel. This option is the fastest one; yet for complex problems, it may lead to a high usage of memory. Therefore, we added an option to choose the number of cores that should be used in the optimization. To provide an indicative idea about the solving times of DIETERpy, we ran a 4-scenario problem with two countries (Germany \& France) for a full year (8760 hours). All additional modules are deactivated, thus only investment into power plant capacities, storage and  net transfer capacities are possible. Between the four scenarios, we only vary the share of renewable energy in both countries (50\%-60\%-70\%-80\% in Germany and 40\%-50\%-60\%-70\% in France). We used a computer that runs Windows 10, has an Intel i5-3320M (2 cores, 4 threads), and 6 GB of RAM. The pure solving times (without any pre- and post-processing of the data) are as follows: sequential solving (no GUSS) 15.0 min, GUSS (sequential) 9.6 min, GUSS (parallel, 4 threads) 8.4 min \footnote{Practitioners can reproduce this example following the instruction in the documentation for example2.}.

GAMS provides the solutions of individual scenario runs in \texttt{GDX} files, which contain all resulting symbols (parameters, variables, equations and sets). When numerous scenario runs have been solved, collecting and combining all symbols can be computationally intensive. Thus, we have designed three solutions: i) the user can select the symbols to be extracted from the \texttt{GDX} files by editing the corresponding file \texttt{reporting\_symbols.csv}; ii) the number of cores can be selected to extract the symbols in parallel; iii) in case many scenarios and symbols are needed, the \textit{VAEX} library consists of streaming algorithms, memory-mapped files and a zero memory copy policy to explore datasets larger than memory \citep{Vaex_2018}. It is advantageous when dealing with resulting multidimensional variables, and processing such variables for several scenarios increases the risk of running out of memory. Hence, \textit{VAEX} reduces the amount of memory needed, although it increases physical storage usage (about 10 times the \textit{GDX} file size). The selected symbols of each scenario run are extracted and saved as separate \texttt{CSV} files or in an individual file per scenario run in case of \texttt{Pickle} and \texttt{VAEX}. As \DIETERpy uses \texttt{Pickle} files to proceed with reporting and visualization, it is the recommended format.


\DIETERpy has been endowed with Python objects that allow the user to make operations between symbols such as addition, subtraction, multiplication and division, while taking care of symbols' dimensions: the \texttt{SymbolsHandler} and \texttt{Symbol} classes help the users to analyze the results more efficiently.

\begin{figure}[b]
    \centering
    \frame{\includegraphics[width=0.99\textwidth]{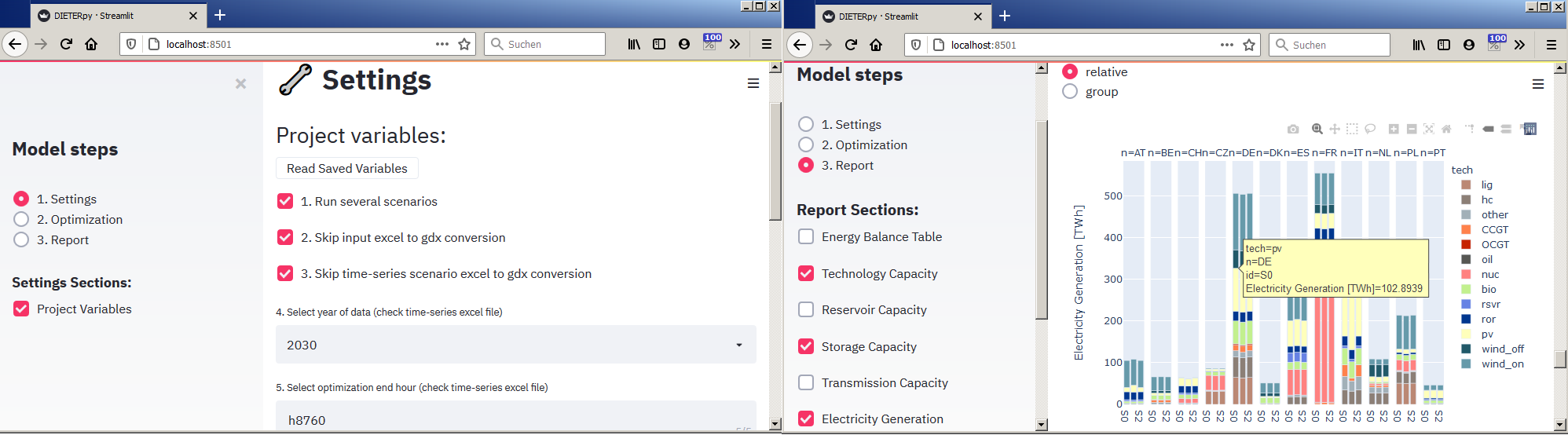}}
    \caption{The graphical user interface. The left panel shows a selection of possible project settings, the right panel model outcomes (source: own illustration).}
    \label{fig:GUI}
\end{figure}

Finally, \DIETERpy provides a browser-based graphical user interface to interactively visualize the results.\footnote{By the time of writing, this interface is under ongoing development and improvement.} Figure \ref{fig:GUI} shows two exemplary screenshots. Additional information on how to use the GUI is provided in the online model documentation.

\section{Illustrative Example} \label{sec: illustrative example}


In this section, we provide an exemplary application of \DIETERpy, varying the costs of stationary Li-ion battery storage in a mid-term future central European setting. In doing so, we highlight some novel Python features introduced by \DIETERpy compared to the previous GAMS-only version \DIETERgms. The example can be reproduced after installing \DIETERpy and creating a project folder by typing in the terminal \texttt{dieterpy create\_project -n $<$project name$>$ -t example1}, where -t stands for \textit{template}.

We adapt \texttt{project\_variables.csv} to change some basic options (see Table \ref{tab:project variables}). For example, we activate the scenario iteration feature (\texttt{scenario\_iteration} set to \texttt{yes}); we decide to run a full ``investment and dispatch model'' (set \texttt{dispatch\_only} is \texttt{no}) with activated electricity exchange between the nodes (\texttt{network\_transfer} set to \texttt{yes}); and we use the maximum number of processor cores for the optimization via the GUSS tool with parallel processes (\texttt{GUSS} is \texttt{yes}, \texttt{GUSS\_parallel} is \texttt{yes} and \texttt{GUSS\_parallel\_threads} is \texttt{0}).

To (de-)activate certain model-related modules, we edit the file \texttt{features\_node\_selection.csv} (Table \ref{tab:Features_node}). In our example, we deactivate all additional modules for simplicity by setting all entries to \texttt{0}. Accordingly, only the basic module is used.

\begin{table}[t]
\centering
\caption{Example of the \texttt{features\_node\_selection.csv} file. Zeros (0) deactivate features, ones (1) activate them. Here, we deactivated all optional features. \texttt{dsm} refers to demand side management, \texttt{ev\_endogenous} and \texttt{ev\_exogenous} to endogenous and exogenous optimization of BEV. For additional information, please refer to the online documentation (source: own illustration)}
\label{tab:Features_node}
\small
\begin{tabular}{lrrrrrrrrrrrr}
\toprule
{} & \multicolumn{12}{c}{Node} \\
Module &   DE & FR & DK & BE & NL & PL & CZ & AT & CH & ES & IT & PT \\
\midrule
dsm           &    0 &  0 &  0 &  0 &  0 &  0 &  0 &  0 &  0 &  0 &  0 &  0 \\
ev\_endogenous &    0 &  0 &  0 &  0 &  0 &  0 &  0 &  0 &  0 &  0 &  0 &  0 \\
ev\_exogenous  &    0 &  0 &  0 &  0 &  0 &  0 &  0 &  0 &  0 &  0 &  0 &  0 \\
reserves      &    0 &  0 &  0 &  0 &  0 &  0 &  0 &  0 &  0 &  0 &  0 &  0 \\
prosumage     &    0 &  0 &  0 &  0 &  0 &  0 &  0 &  0 &  0 &  0 &  0 &  0 \\
heat          &    0 &  0 &  0 &  0 &  0 &  0 &  0 &  0 &  0 &  0 &  0 &  0 \\
\bottomrule
\end{tabular}
\end{table}

\begin{table}[b]
\centering
\caption{Example of the \texttt{iteration\_table.csv} file with different assumptions on the annualized costs of Li-ion storage (source: own illustration).}
\label{tab:main iteration}
\begin{tabular}{lrr}
\toprule
{} &  c\_i\_sto\_e(n,'Li-ion') &  c\_i\_sto\_p(n,'Li-ion') \\
run &  [EUR/MWh]              &     [EUR/MW]               \\
\midrule
S0   &                  20029 &                  15021 \\
S1   &                  10014 &                   7511 \\
S2   &                   5007 &                   3755 \\
\bottomrule
\end{tabular}
\end{table}

We analyze the effects of lower-cost Li-ion storage with three different scenario runs. Model run S0 represents baseline storage cost assumptions; the scenario S1 reflects a cost decrease of both energy- and power-related Li-ion investments of 50\%; and scenario S2 assumes a further storage cost reduction to 25\% of S0 values. We specify Table \ref{tab:main iteration} accordingly. \DIETERgms defines these parameters as \texttt{c\_i\_sto\_e(n,sto)} and \texttt{c\_i\_sto\_p(n,sto)} for energy and power components, respectively. First, as we do not provide the special features \texttt{country\_set} as a column heading, we choose the entire set of \textit{n} countries to be included in the runs. Second, the set \textit{sto} in the model consists of three elements: ``Li-ion'' (stationary battery storage), ``PHS'' (pumped hydro storage, which is fixed in this example) and ``P2G2P'' (power-to-gas-to-power). As we want to vary only the costs of Li-ion batteries, we provide it literally in the column heading: \texttt{c\_i\_sto\_e(n,`Li-ion')} and \texttt{c\_i\_sto\_p(n,`Li-ion')}.

\begin{figure}[b!]
    \centering
    \frame{\includegraphics[width=0.95\textwidth]{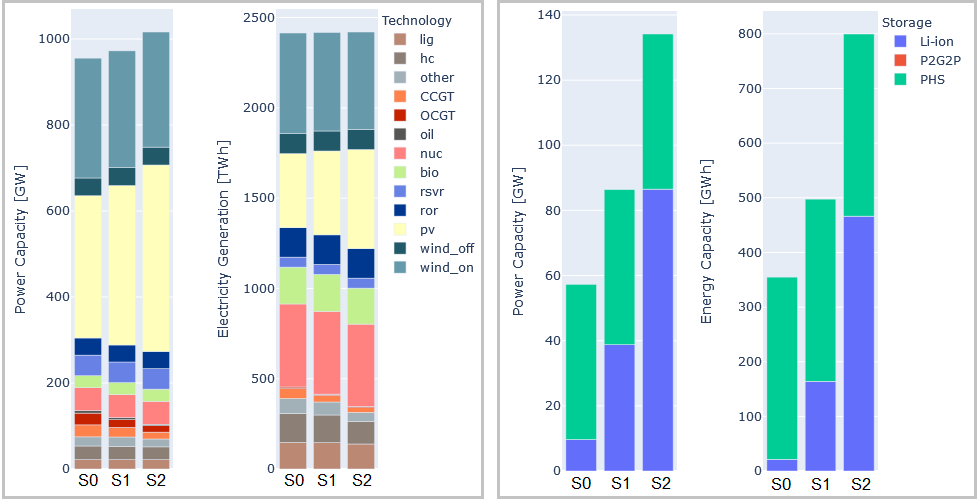}}
    \caption{Installed capacity and total generation of conventional and renewable generators (left panel) and storage (right panel) for three exemplary scenarios (source: own illustration).}
    \label{fig:results capacities}
\end{figure}

Here we only present a snapshot of the results of the exemplary application, using figures generated by our built-in visualization tool. Cheaper batteries generally increase the use of solar PV, both regarding installed capacity and yearly electricity generation (left panel of Figure \ref{fig:results capacities}). This is because short-term battery storage is particularly suited to balance daily PV fluctuations. If the costs of Li-ion batteries decrease as assumed here, they would be deployed with comparable energy and power capacity as existing European pumped hydro storage (right panel of Figure \ref{fig:results capacities}).

\begin{figure}[t]
    \centering
    \frame{\includegraphics[width=0.95\textwidth]{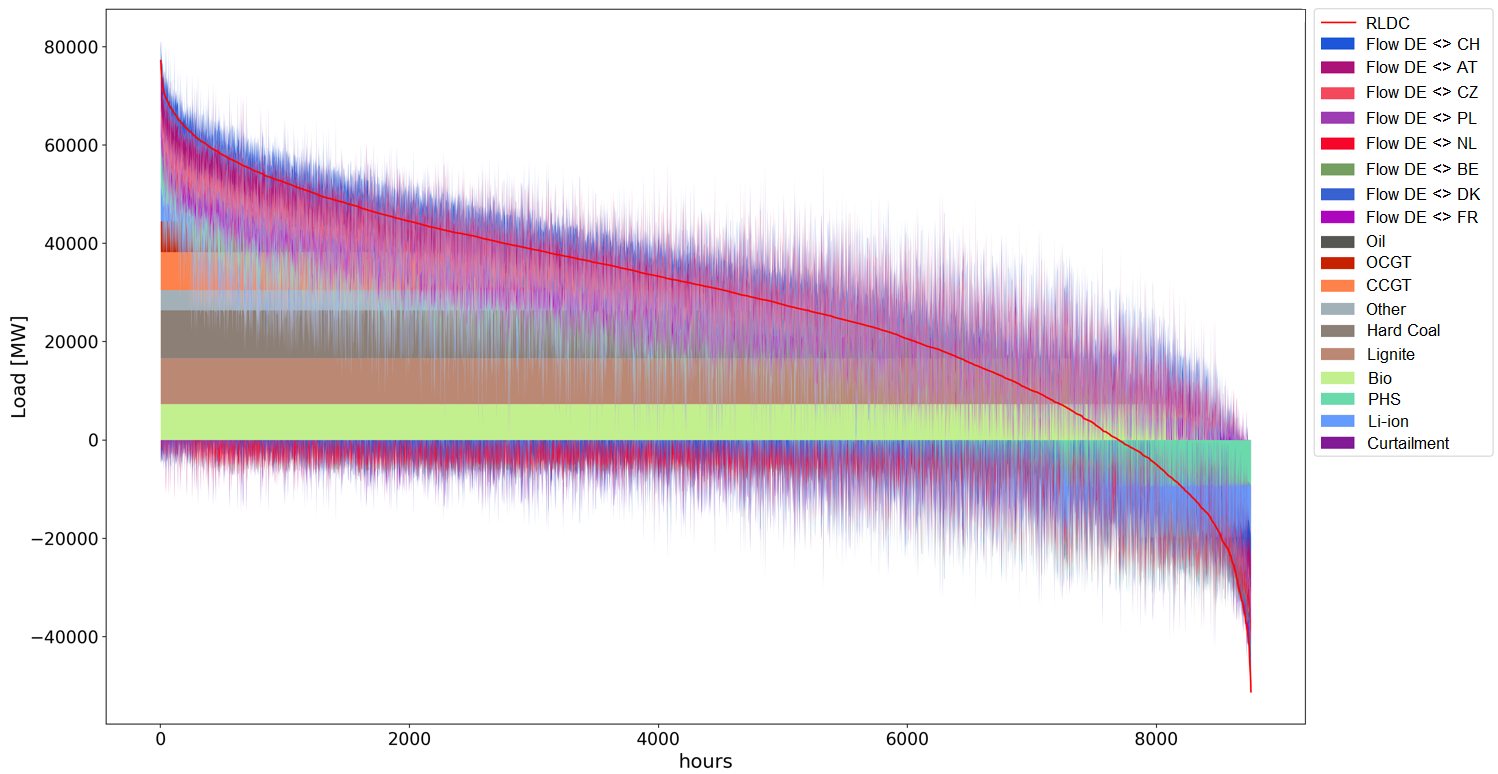}}
    \caption{Residual load duration curve, generation of dispatchable technologies and storage, and transmission flows. This example represents scenario S2 for Germany (source: own illustration).}
    \label{fig:results exemplary rldc DE}
\end{figure}

Figure \ref{fig:results exemplary rldc DE} shows an exemplary residual load duration curve (RLDC), combined with the generation of various technologies in respective hours. The typical role of different generation technologies can be seen, e.g.,~base-load use of lignite and peak-load use of open cycle gas turbines (OCGT). Storage charging mainly occurs in periods of renewable surplus generation, but partly also in other periods.

\section{Impact \& Conclusion}\label{sec:impact}



DIETER has been designed to investigate research questions that are highly relevant in current energy transition research. This includes, for example, the exploration of infrastructure requirements for least-cost integration of variable renewable energy sources in the power sector, and the effects of flexibility using wind and solar energy also in other sectors. The model's prosumage module further allow investigating specific aspects of decentralized self-supply with PV-battery systems, which is usually not covered by other energy models.

The new features of the \DIETERpy model version presented here will increase the efficiency of numerical model analyses applied to respective research questions. Compared to previous model versions, the new Python implementation allows to more easily specify, run, and evaluate the outcomes of numerous scenarios, and to make use of a great variety of Python libraries for data pre- and post-processing and for visualizing results.

    
Despite the new Python functionalities, \DIETERpy in its core remains a parsimonious model which can be flexibly applied to different parametrizations and research questions. Most of the applications we published so far could be solved on standard desktop computers. Only very large-scale applications covering many regions and multiple sector coupling options may require larger computational resources. Accordingly, \DIETERpy may not only be used by highly specialized research groups, but also by students and practitioners.
    
The Python framework presented here further allows model users without deeper knowledge of programming in either GAMS or Python to run a large range of model applications. In particular, the browser-based graphical user interface increases the ease of use and the accessibility of the model. More expert users are free to adjust and expand any part of \DIETERpy as they wish, given the open-source nature of the model and the permissive MIT license.


Thus, we expect that the \DIETERpy version presented here will contribute to the further usage and application of the model, especially outside the domain of the current development team. In the past, most publications based on DIETER were co-authored by DIW Berlin researchers. Two articles introduce the basic model version and investigate optimal electrical storage capacity in scenarios with high shares of renewable energy sources \cite{ZERRAHN_2017,SCHILL_2018}. Reduced model versions are used for more general reflections of the economics of electrical storage \cite{ZERRAHN_2018} and its changing role in settings with increasing renewable penetration \cite{Schill_Joule_2020}. Three papers on solar prosumage focus on power sector effects for Germany \cite{Schill_2017} and Western Australia \cite{say2020}, and on how tariff design impacts PV-battery investments \cite{guenther_2019}. Further model applications analyze the power sector impacts of electric vehicles \cite{schill_2016}, flexible electric heating \cite{Schill_2020}, and green hydrogen \cite{stoeckl2020}. A demand-side management feature we developed for DIETER was published separately \cite{zerrahn_2015}. Recently. we also noticed spin-off versions of the model that have been developed without any involvement of the initial DIETER team. These include, for example, generation cost projections developed for the Australian Energy Market Operator \cite{graham_gencost_2018}, as well as analyses of demand-side management in India \cite{Ershad_2020} and methodological aspects of modeling long-term storage \cite{deguibert_2020}. We expect that the \DIETERpy improvements described here will spur more of such activities.
    

    
Because of the permissive license, \DIETERpy can also be used in commercial settings. Yet we assume that it is more likely to be used in academic and not-for-profit research applications. In fact, the model and its applications have been the backbone of several research grants acquired by us, including such of different German federal ministries and the European Commission.

To draw a more general conclusion, with \DIETERpy we show how to embed an existing GAMS-based model into a modern Python framework. Now, the model combines the best of the two programming languages. On the one hand, GAMS is a well-established language for energy models which is proofed to work. It offers a straightforward algebraic formulation and the use of efficient solvers. Maintaining the GAMS core also allows easier linking with other GAMS-based models. On the other hand, Python offers more flexibility, a wide choice of additional libraries, and better opportunities for data pre- and post-processing as well as visualization of results. Using Python also enables an easy-to-use and flexible scenario tool which would be difficult to implement in GAMS. This gives the user the opportunity to run the model in many different scenarios by flexibly altering the model parametrization or its constraints. We believe that \DIETERpy could serve as an example also for other established energy models to make them future-proof and more accessible to a wider audience.

Several model improvements and upgrades are currently ongoing or planned. This concerns both the algebraic GAMS core, where we aim for a more complete and more consistent coverage of different sector coupling options, and further advancements of the Python framework. We particularly aim to improve and extended the user interface and the post-processing and data visualization functionalities.

\section{Conflict of Interest}\label{sec:conflict of interest}



No conflict of interest exists: we confirm that there are no known conflicts of interest associated with this work.

\section*{Acknowledgments} \label{sec:acknowledgements}

This work has been supported by research grants of the German Federal Ministry of Education and Research via the projects ``Future of Fossil Fuels in the wake of greenhouse gas neutrality'', FKZ 01LA1810B, and ``Ariadne'', FKZ 03SFK5NO.

\section*{Author Contributions} \label{sec:author contributions}


\textbf{Carlos Gaete:} Conceptualization; Methodology; Software (Python); Visualization; Writing - original draft, review and editing. \textbf{Martin Kittel:} Methodology; Software (Python); Writing - review and editing. \textbf{Alexander Roth:} Methodology; Software (Python); Writing - original draft, review and editing; Online documentation. \textbf{Wolf-Peter Schill:} Funding acquisition; Methodology; Software (GAMS); Writing - original draft, review and editing.





\bibliographystyle{elsarticle-num}




\end{document}